\documentclass[12pt]{iopart}
\usepackage{graphicx}
\usepackage{epsfig}

\begin{document}

\title[Schr\"{o}dinger-Newton ``collapse" of the wave function]
{Schr\"{o}dinger-Newton ``collapse" of the wave function}

\author{J R van Meter}
\address{Dept. of Physics, University of Maryland, Baltimore County, 1000 Hilltop Circle, Baltimore, MD 21250, USA}
\ead{james.r.vanmeter@nasa.gov}

\begin{abstract}
It has been suggested that the nonlinear Schr\"odinger-Newton equation might approximate the coupling of quantum mechanics with gravitation, particularly in the context of the M{\o}ller-Rosenfeld semiclassical theory.
Numerical results for the spherically symmetric, time-dependent, single-particle case are presented, clarifying and extending previous work on the subject.  It is found that, for a particle mass greater than $1.14(\hbar^2/(G\sigma))^{1/3}$, a wave packet of width $\sigma$ partially ``collapses" to a groundstate solution found by Moroz, Penrose, and Tod, with excess probability dispersing away.  However, for a mass less than $1.14(\hbar^2/(G\sigma))^{1/3}$, the entire wave packet appears to spread like a free particle, albeit more slowly.  
It is argued that, on some scales (lower than the Planck scale), this theory predicts significant deviation from conventional (linear) quantum mechanics.  However, owing to the difficulty of controlling quantum coherence on the one hand, and the weakness of gravity on the other, definitive experimental falsification poses a technologically formidable challenge.
\end{abstract}

\section{Introduction}
\label{Introduction}

Noting there is no direct evidence that the gravitational field is quantized,
Carlip has suggested the possibility that it is not \cite{Carlipb}.  The suggestion is not new, and Carlip points specifically to a theory put forth by Rosenfeld \cite{Rosenfeld} and independently
by M{\o}ller \cite{Moller}, that the gravitational field retains its classical nature while coupling to
an expectation value of quantized matter:
\begin{equation}
G_{ab} = 8\pi \langle T_{ab} \rangle.
\label{a2}
\end{equation}
Although a superficially similar equation has been discussed in the context of a semiclassical approximation to quantum gravity \cite{Wald}, M{\o}ller and Rosenfeld proposed that Eq.~(\ref{a2}) is exact.  Despite its unconventional character, Carlip has argued that the theory has not yet been ruled out, and with Salzman has worked towards an experimentally testable prediction \cite{Salzman,Carlipa,Carlipb}.  The intent of the present work is to continue this effort with more accurate numerical results.


In the nonrelativistic limit, $\langle T_{ab}\rangle=\langle\Psi|\hat{T}_{ab}|\Psi\rangle$ where $\Psi$ is a state function that evolves according to Schr\"odinger's equation \cite{Kibble,Diosi},  
\begin{equation}
i\hbar\frac{\partial\Psi(\mathbf{X},t)}{\partial t} = -\sum_{i=1}^N\frac{\hbar^2}{2m_i}\nabla_i^2\Psi(\mathbf{X},t) - G\sum_{i=1}^Nm_i\Phi(x_i,t)\Psi(\mathbf{X},t),
\label{schrodinger}
\end{equation}
and the potential $\Phi$ is given by the Newtonian limit of Eq.~(\ref{a2}), assuming $N$ source particles,
\begin{equation}
\nabla^2\Phi(x,t)=4\pi G\int d^{3N}X'|\Psi(\mathbf{X}',t)|^2\sum_{j=1}^Nm_j\delta(\mathbf{x}-\mathbf{x}_j'),
\label{manyn}
\end{equation}
and $\mathbf{X}\equiv (\mathbf{x}_1,\ldots,\mathbf{x}_N)$.
Eq.~(\ref{schrodinger}) coupled with Eq.~(\ref{manyn}) is known as the Schr\"odinger-Newton system,
which can be written as a single equation by solving Eq.~(\ref{manyn}) with the Green's function:
\begin{equation}
i\hbar\frac{\partial\Psi(\mathbf{X},t)}{\partial t} = -\sum_{i=1}^N\frac{\hbar^2}{2m_i}\nabla_i^2\Psi(\mathbf{X},t) - G\sum_{i,j=1}^Nm_im_j\int d^{3N}X'\frac{|\Psi(\mathbf{X}',t)|^2}{|\mathbf{x}_i-\mathbf{x}_j'|}\Psi(\mathbf{X},t)
\label{manysn}
\end{equation}

Conventionally, a Schr\"odinger potential excludes nonlinear self-interaction.  In violating this rule, Eq.~(\ref{manysn}) 
represents a fundamental alteration of quantum mechanics.  First, because it includes self-interaction, the potential is present for any massive particle irrespective of the presence of other particles or any external potential.
There are no free particles in this theory.  Second, it is nonlinear even for a single particle.  Thus, arbitrary linear superpositions are no longer 
permitted (although they will often serve as good approximations when the mass is small enough that the nonlinear term can be neglected).

It should be emphasized that this Schr\"odinger-Newton system differs markedly from the conventional coupling of
the Schr\"odinger equation with a Newtonian or Coulomb-like potential, 
as noted in \cite{Adler}.
It is true that the Schr\"odinger-Newton equation shares terms in common with the standard mean-field
approximation of such a potential.  For example, substituting the Hartree ansatz $\Psi=\psi(\mathbf{x}_1,t)\cdots\psi(\mathbf{x}_N,t)$  into the conventional system gives:
\begin{eqnarray}
i\hbar\frac{\partial\psi_i(\mathbf{x}_i,t)}{\partial t} = &-&\frac{\hbar^2}{2m_i}\nabla_i^2\psi
_i(\mathbf{x}_i,t) -  G\sum_{j=1}^Nm_im_j\int d^{3}x_j'\frac{|\psi_j(\mathbf{x}_j',t)|^
2}{|\mathbf{x}_i-\mathbf{x}_j'|}\psi_i(\mathbf{x}_i,t)  \nonumber\\
&+& Gm_i^2\int d^{3}x_i'\frac{|\psi_i(\mathbf{x}_i',t)|^2}{|\mathbf{x}_i-\mathbf{x}_i'|}\psi_i(\mathbf{x}_i,t)
\label{hartree}
\end{eqnarray}
(Usually the sum in Eq.~(\ref{hartree}) excludes the $j=i$ term; here the $j=i$ term is included in the sum while its negation is separately added, for emphasis.)
Applying the same ansatz to the Schr\"odinger-Newton system yields:
\begin{eqnarray}
i\hbar\frac{\partial\psi_i(\mathbf{x}_i,t)}{\partial t} = -\frac{\hbar^2}{2m_i}\nabla_i^2\psi_i(\mathbf{x}_i,t) &-&  G\sum_{j=1}^Nm_im_j\int d^{3}x_j'\frac{|\psi_j(\mathbf{x}_j',t)|^2}{|\mathbf{x}_i-\mathbf{x}_j'|}\psi_i(\mathbf{x}_i,t) 
\label{hartreesn}
\end{eqnarray}
These equations are similar but Eq.~(\ref{hartreesn}) includes self-interaction, whereas Eq.~(\ref{hartree}) excludes it.
This distinction is also true of the Hartree-Fock approach, with an ansatz consisting of the Slater determinant and a potential corrected for fermion statistics.
In this case the final term in Eq.~(\ref{hartree}) is replaced by a sum called the ``exchange term" which, again, cancels the self-interaction term in the conventional case.

This distinguishing feature is most obvious for a single-particle system.  In this case, the standard mean-field potential vanishes, leaving the equation for a free particle.
Meanwhile the Schr\"odinger-Newton system retains the self-interaction term:
\begin{equation}
i\hbar\frac{\partial\psi(\mathbf{x},t)}{\partial t} = -\frac{\hbar^2}{2m}\nabla^2\psi(\mathbf{x},t) - Gm^2\int d^3x'\frac{|\psi(\mathbf{x}',t)|^2}{|\mathbf{x}-\mathbf{x}'|}\psi(\mathbf{x},t).
\label{singlesn}
\end{equation}

In the many-particle case, the significant physical difference between the Schr\"odinger-Newton and conventional Hartree or Hartree-Fock
systems is made more evident by transforming to relative and center-of-mass coordinates,
$\mathbf{R} \equiv \sum_{i=1}^N m_i\mathbf{x}_i/M$, $M\equiv\sum_{i=1}^Nm_i$, $\mathbf{r}_j \equiv \mathbf{x}_j-\mathbf{x}_N$, $j=1,\ldots,N-1$,
and using the ansatz $\Psi=\psi_R(\mathbf{R},t)\psi_{r_1}(\mathbf{r}_1)\cdots\psi_{r_{N-1}}(\mathbf{r}_{N-1},t)$ (or the Slater determinant).
In the conventional system, the potential can be expressed entirely in terms of relative 
coordinates.  Separation of variables then leaves the center-of-mass equation identical in form to that of a free-particle with mass $M$ and position $\mathbf{R}$. 
In the Schr\"odinger-Newton system, on the other hand, 
the potential cannot be expressed entirely in terms of relative coordinates,
because of the appearance of the kernel
$|\mathbf{x}_N-\mathbf{x}'_N|^{-1}$.
In this case the center-of-mass receives self-interaction,
and separation of variables results in a center-of-mass equation identical in form to the nonlinear, single-particle Schr\"odinger-Newton equation (Eq.~(\ref{singlesn})), with mass $M$ and position $\mathbf{R}$.

There have been various efforts to investigate physical consequences of this nonlinearity.
Diosi argued that it would result in soliton-like behavior, 
which he predicted would effectively localize macroscopic objects \cite{Diosi}.  He further estimated the width and energy of the single-particle groundstate.  Independently Penrose conjectured that quantum gravity requires gravitationally induced collapse of the wave function, and suggested the Schr\"odinger-Newton equation might approximate this process \cite{Penrose96,Penrose98,Moroz}.
This motivated a numerical study by Moroz, Penrose, and Tod (MPT), which yielded stationary solutions in the single particle case,
including a groundstate consistent with Diosi's estimate \cite{Moroz}.

Subsequent numerical studies focused on the time-dependent, single-particle Schr\"odinger-Newton equation.
The following were among the features observed of the solution set:
\begin{itemize}
\item Some solutions tended to contract in radius, for the most part evolving towards a narrower wave function.  This was observed by Harrison et al.\cite{Harrison01,Harrison02,Harrison}, Salzman and Carlip \cite{Salzman,Carlipa,Carlipb}, and Giulini and Grossardt \cite{Giulini}.  Guzman et al. also made similar observations of Schr\"odinger-Newton solutions albeit in a different physical context, that of hypothetical scalar fields in astrophysical systems \cite{Guzman04,Guzman06,Bernal}.  Harrison et al. identified the narrower wave function  as a scaled-down MPT groundstate.
\item These same solutions fragmented slightly, releasing pulses carrying a small fraction of their total probability to infinity, while the rest of their amplitude remained more localized.  This was observed by Harrison et al. and  Guzman et al.  
\item Below a critical mass, solutions were found which seemed to expand, in their entirety, without bound, much like free particles but with slower velocities.  This was observed by Salzman and Carlip, and Giulini and Grossardt, although the critical mass found by the latter was six orders of magnitude greater than that found by the former.
\end{itemize}

The present study qualitatively demonstrates all of the above features.  Quantitatively, the results presented here are consistent with Harrison et al. in regards to the final state towards which some solutions evolve, to within a few percent, and with Giulini and Grossardt in regards to the value of the critical mass below which solutions expand towards spatial infinity, to within $\sim 20\%$.  (Note the present work was originally submitted before the Giulini and Grossardt preprint appeared.)

We remark in passing that the term ``collapse", which is used to describe aspects of Schr\"odinger-Newton evolution here and in \cite{Salzman,Carlipa,Carlipb}, should be qualified.  The term is used because in some respects the dynamics resemble the traditional conception of collapse of the wave function: for an amount of matter approaching macroscopic scales, the wave function tends to localize
in position and approach an energy eigenstate.  However, unlike some traditional
notions about wave function collapse, Schr\"odinger-Newton evolution is continuous and unitary. 

The present investigation will focus on the numerical evolution of a single particle with the same initial data used by Salzman and Carlip:
\begin{equation}
\psi(r,0)=(\pi\sigma^2)^{-3/4}e^{-r^2/(2\sigma^2)}
\label{initialpsi}
\end{equation}
This paper is organized as follows:  
In Sec. 2 the numerical methods are explained.  In Sec. 3 the main numerical results are given.  And in Sec. 4 physical implications and the possibility for empirical verification are discussed.

\section{Methods}
\label{Methods}

As in \cite{Salzman} for a single-particle source and spherical symmetry the Poisson equation (Eq.~(\ref{manyn})) can be solved using the 
appropriate Green's function,
\begin{equation}
G(\vec{r},\vec{r}')=\frac{1}{4\pi\max(r,r')},
\end{equation}
yielding
\begin{eqnarray}
\Phi&=-&\int_0^{\infty}G(\vec{r},\vec{r}')|\psi(\vec{r}',t)|^2 dV'\nonumber\\
&=& -4\pi Gm \left[ r^{-1}\int_0^{r'} r^{'2} |\psi(r',t)|^2 dr'+\int_r^{\infty} r' |\psi(r',t)|^2 dr'\right].
\label{phi}
\end{eqnarray}
In the present work this is integrated with a 4th order accurate Newton-Cotes algorithm (i.e. the error is $\mathcal O(\Delta r^4)$ where $\Delta r$ is the interval between adjacent points of the discretized radius).  
\begin{figure} [h]
\includegraphics*[width=25pc,height=40pc,angle=-90]{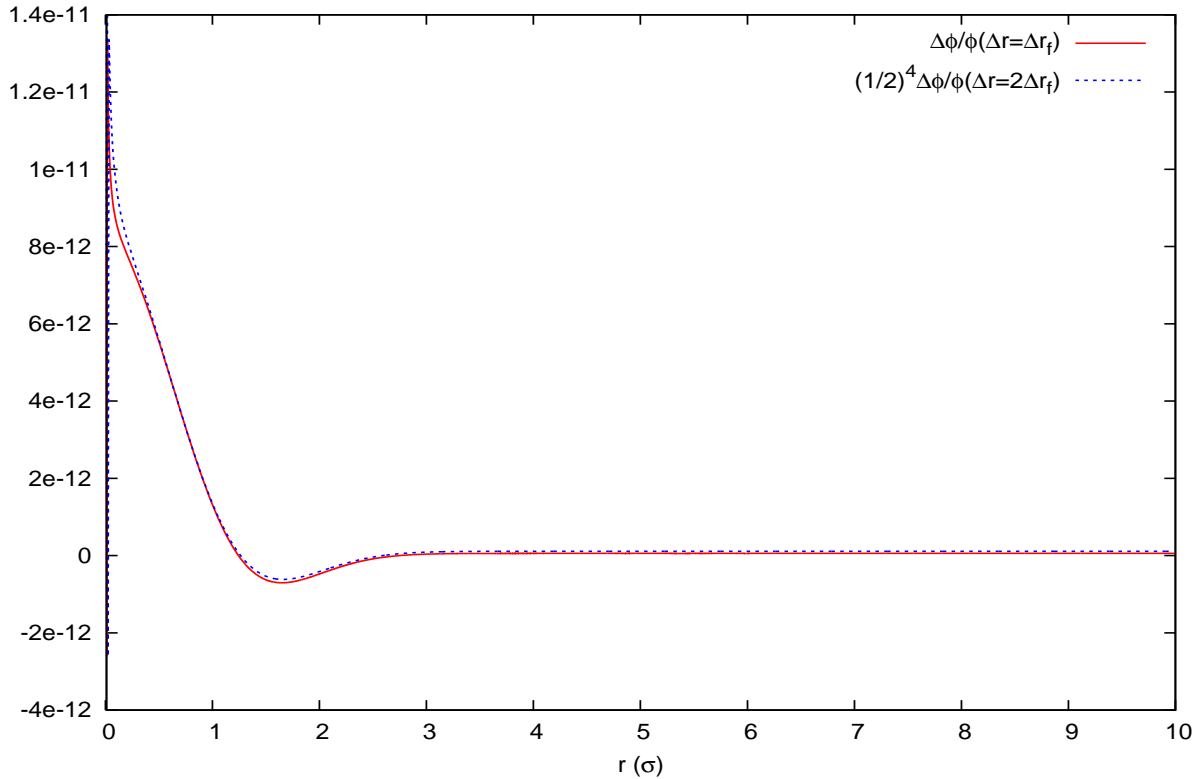}
\caption{Relative error for the initial computation of $\Phi$ is shown, at two different resolutions.  The scaling is such that for the expected 4th order accuracy the two curves should coincide.}
\label{fig:phiconv}
\end{figure}
Fig.~\ref{fig:phiconv} demonstrates its accuracy by comparing its result for the initial $\Phi$ to the analytic solution, $Gm^2\mbox{erf}(r/\sigma)/r$, obtained from Eq.~(\ref{phi}) with the initial $\psi$ of Eq.~(\ref{initialpsi}).  

As in \cite{Harrison}, Schr\"{o}dinger's equation can be simplified slightly in spherical symmetry by the change of variables $u=r\psi$:  
\begin{equation}
\partial_t u = i\frac{\hbar}{2m}\partial_r^2u - i\frac{m}{\hbar}\Phi u
\label{simple}
\end{equation}
Here the spatial derivative on the right hand side is computed with a 6th order accurate stencil.  To provide a boundary condition appropriate for finite differencing near the
origin, $u(r)$ is assumed to be effectively an odd function (as is consistent with the initial data and Eq.~(\ref{simple})).
  Time-integration is then performed with a 4th order accurate Runge-Kutta algorithm (that is, 4th order in $\Delta t$).  Note that stability requires that 
$\Delta t\propto \Delta r^2$
Thus the time-integration is 8th order in $\Delta r$, and the dominant error,
outside of the Poisson-integration, should be 6th order in $\Delta r$.  
\begin{figure} [h]
\includegraphics*[width=25pc,height=40pc,angle=-90]{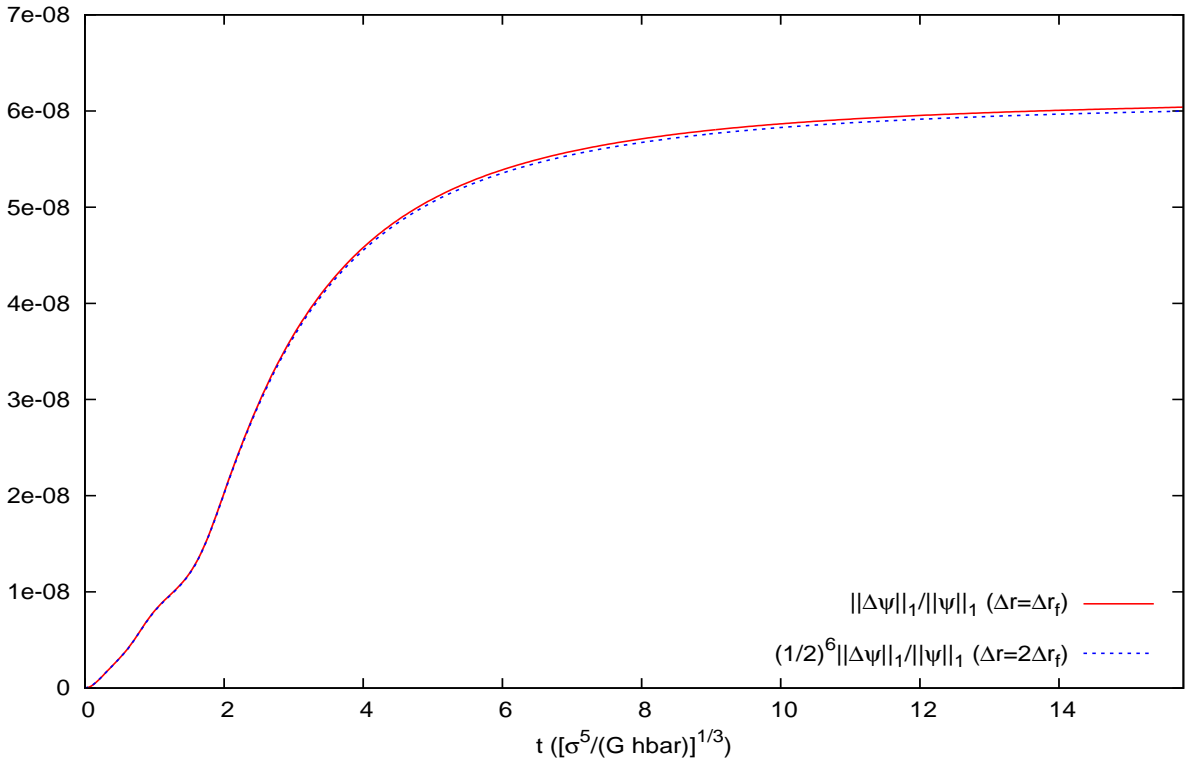}
\caption{The $L_1$ norm of the relative error of a free particle wave function is shown, at two different resolutions, as a function of time.  The scaling is such that for the expected 6th order accuracy the two curves should coincide.}
\label{fig:psiconv}
\end{figure}
Fig.~\ref{fig:psiconv} shows this is indeed the case, by comparing the results of a simulation with zero potential with the analytic free-particle solution.

It will prove convenient to express units in terms of $G$, $\hbar$, and $\sigma$.  The basic unit of distance is then $\sigma$, the basic unit of time is $(\sigma^5/(G\hbar))^{1/3}$,
and the basic unit of mass is $(\hbar^2/(G\sigma))^{1/3}$.  
These units exhibit the scale-invariance of the Schr\"{o}dinger-Newton equation: a given solution remains a solution if the distance is multiplied by $\lambda$, time is multiplied by $\lambda^{5/3}$, 
and mass is multiplied by $\lambda^{-1/3}$, (and $\psi$ is renormalized accordingly).
For an example of a particular physical scale, if $\sigma=4.47\times 10^{-9}$m, as in \cite{Carlipb}, then time is given in units of 6.33 seconds and mass is given in units of $3.33\times 10^{-17}$kg ($2\times 10^{10}$ u).

Since the computational domain must be finite, the outer boundary is located at an arbitrary position but far enough away that 
$|\psi|$ remains negligible there throughout most of the evolution.  
The boundary condition is chosen to be 
Neumann.  
This choice is arbitrary, due to the following machination.
As in \cite{Becerril}, in order suppress reflections, the potential is modified with the addition of 
an imaginary $\tanh$ term, constructed so as to be of order 1 within  $\sim\sigma$ of the outer boundary.
This term is only non-negligible near the outer boundary, where it dampens waves of wavelength less than $\sim\sigma$, effectively simulating their propagation beyond the computational domain.

Some additional nontrivial computations are performed for purposes of analysis. The expectation value of the Hamiltonian, $\langle\hat{H}\rangle =\langle \psi|\hat{H}|\psi\rangle$, is computed in order to show its approach to an eigenenergy as $\psi$ approaches a stationary state,
with the formula,
\begin{equation}
\langle\hat{H}\rangle =\int \left[ -\frac{\hbar^2}{2m}\psi^* \nabla^2\psi+m\Phi|\psi|^2 \right] dV
\end{equation}
where the differentiation and integration is performed numerically as described above.
In addition, it is useful to find the peak of $r^2|\psi|^2$.  This is done by locating, among the discrete radii, the point at which $r^2|\psi|^2$ is maximum, then 
fitting this and the neighboring points to a polynomial.

\section{Results}
\label{Results}

For $m > 1.14(\hbar^2/(G\sigma))^{1/3}$ the wave packet generally fragmented in two.  A fraction of the probability, with wavelength of order $\hbar^2/(Gm^3)$, appeared to propagate outwards to infinity.
The rest remained mostly localized in a finite region.  
\begin{figure} [h]
\includegraphics*[width=25pc,height=40pc,angle=-90]{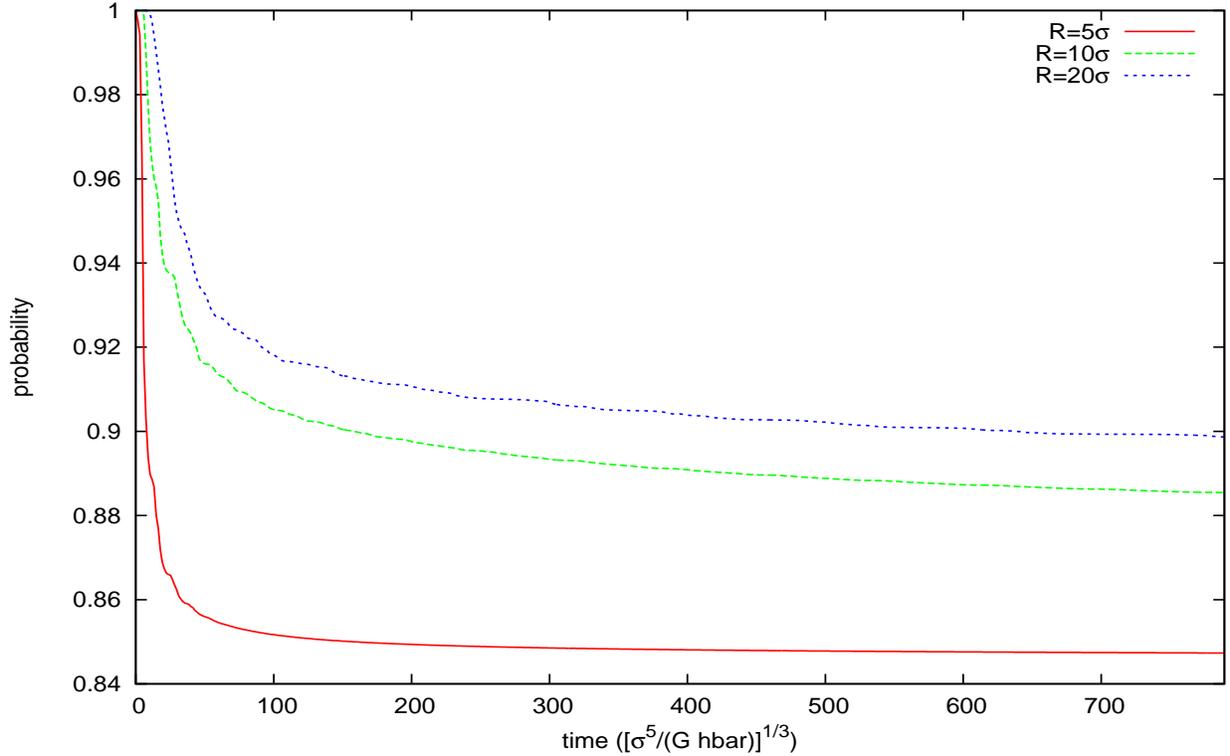}
\caption{Total probability in the computational domain is shown, as a function of time, for three different outer boundary locations $R$.  Most of the probability is clearly localized within a finite region, and asymptotes to a constant.}
\label{fig:norm}
\end{figure}
Fig.~\ref{fig:norm} shows the amount of probability remaining in the computational domain during the course of a simulation, for $m=2.09(\hbar^2/(G\sigma))^{1/3}$ and three
different domain sizes.  It is apparent that most of the probability is highly localized, and constant.

\begin{figure} [h]
\includegraphics*[width=25pc,height=40pc,angle=-90]{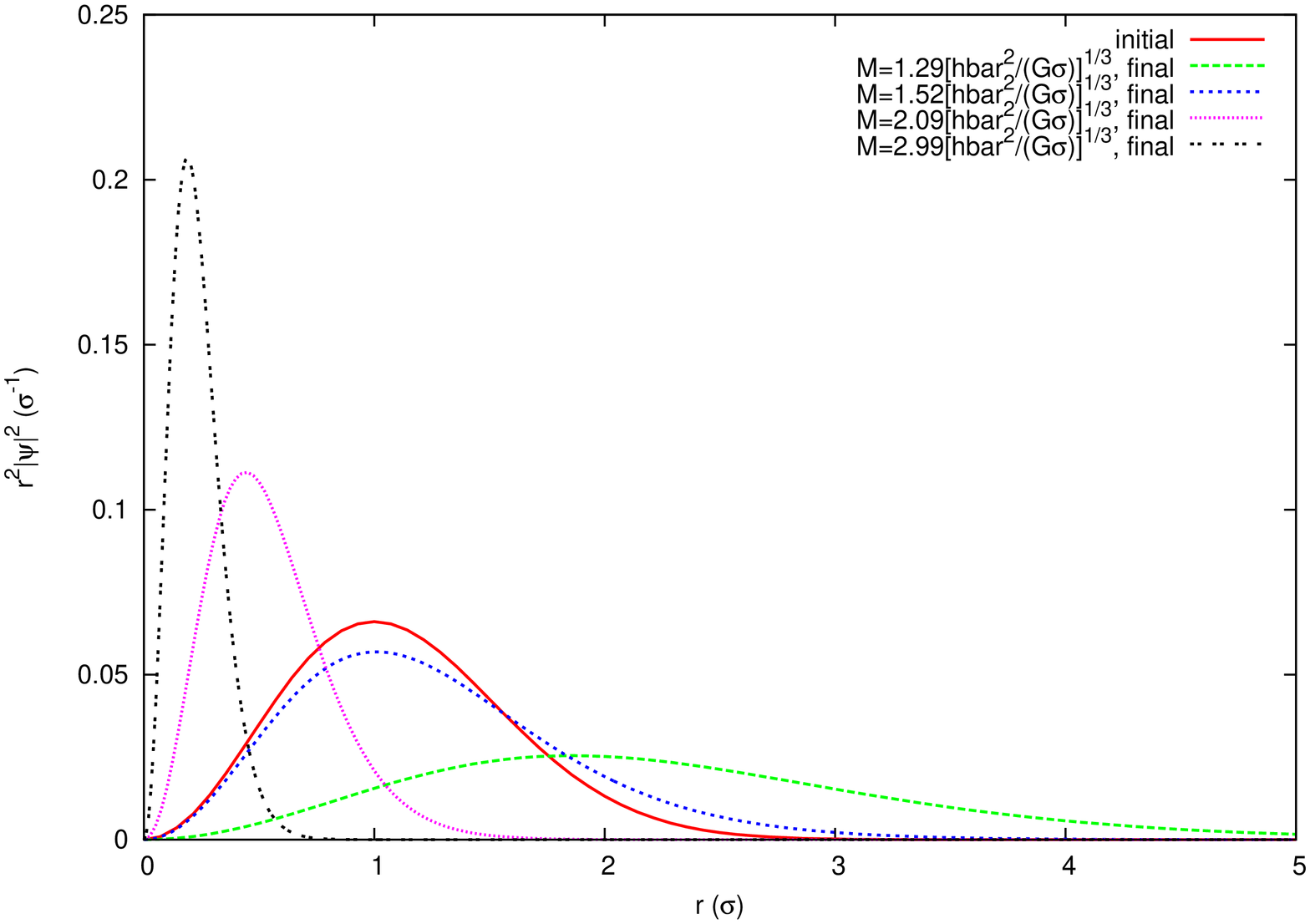}
\caption{The probability density shown late in evolution for four different masses, all starting from the same initial state, also shown.  Each of the ``final" states 	is approximately stationary, oscillating only slightly about these amplitudes.  The radius of the peak scales as $\hbar^2/(Gm^3)$. }
\label{fig:r2psi2}
\end{figure}

\begin{figure} [h]
\includegraphics*[width=25pc,height=40pc,angle=-90]{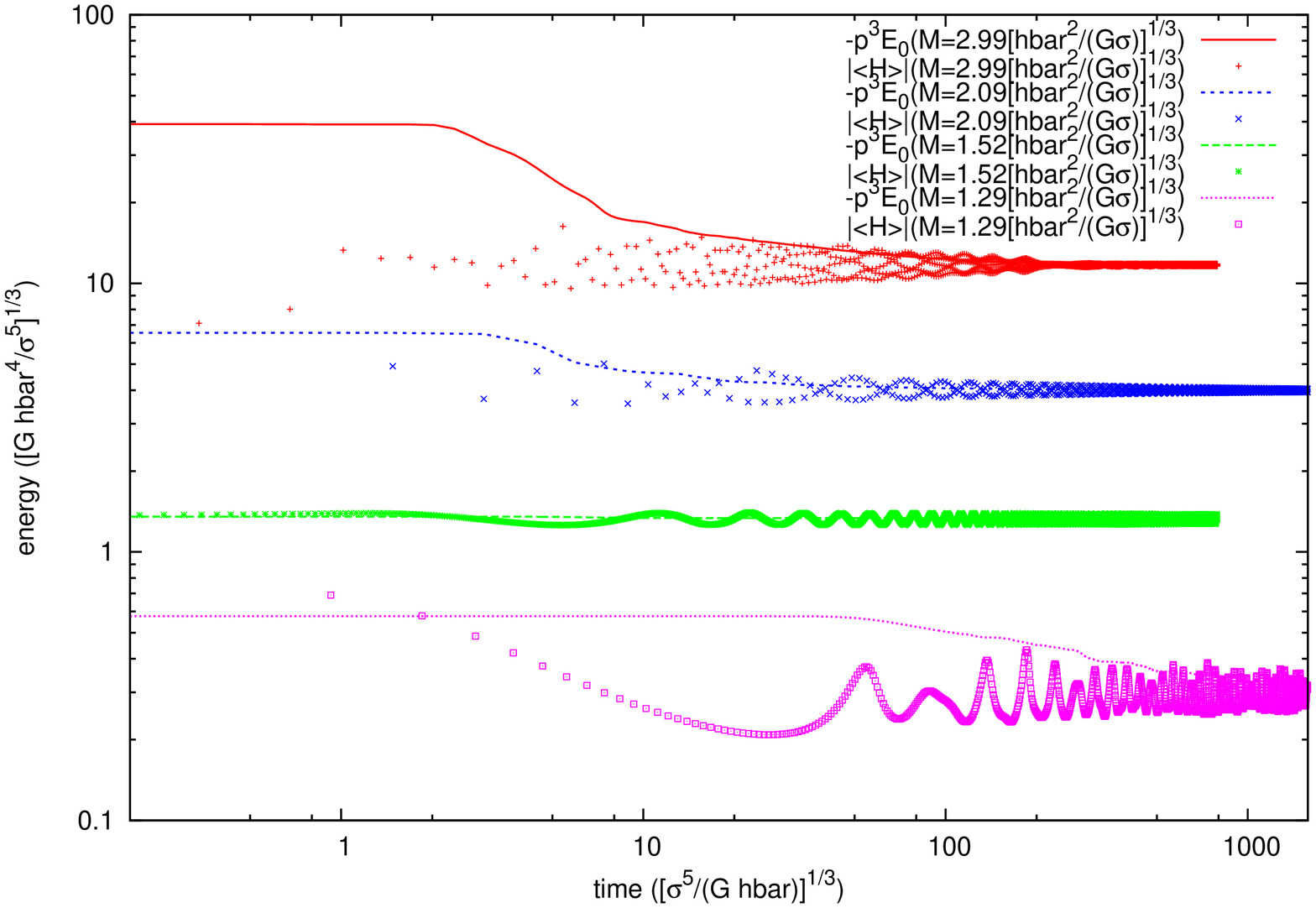}
\caption{$\langle\hat{H}\rangle$ computed during the evolution of four masses.  Each approaches the groundstate energy 
$E_0=-0.163G^2m^5/\hbar^2$, 
rescaled by the cube of the probability.
}
\label{fig:energy}
\end{figure}

This remnant wave function appeared to eventually approach a stationary state, about which it performed damped oscillations.
\begin{figure}
\includegraphics*[width=25pc,height=40pc,angle=-90]{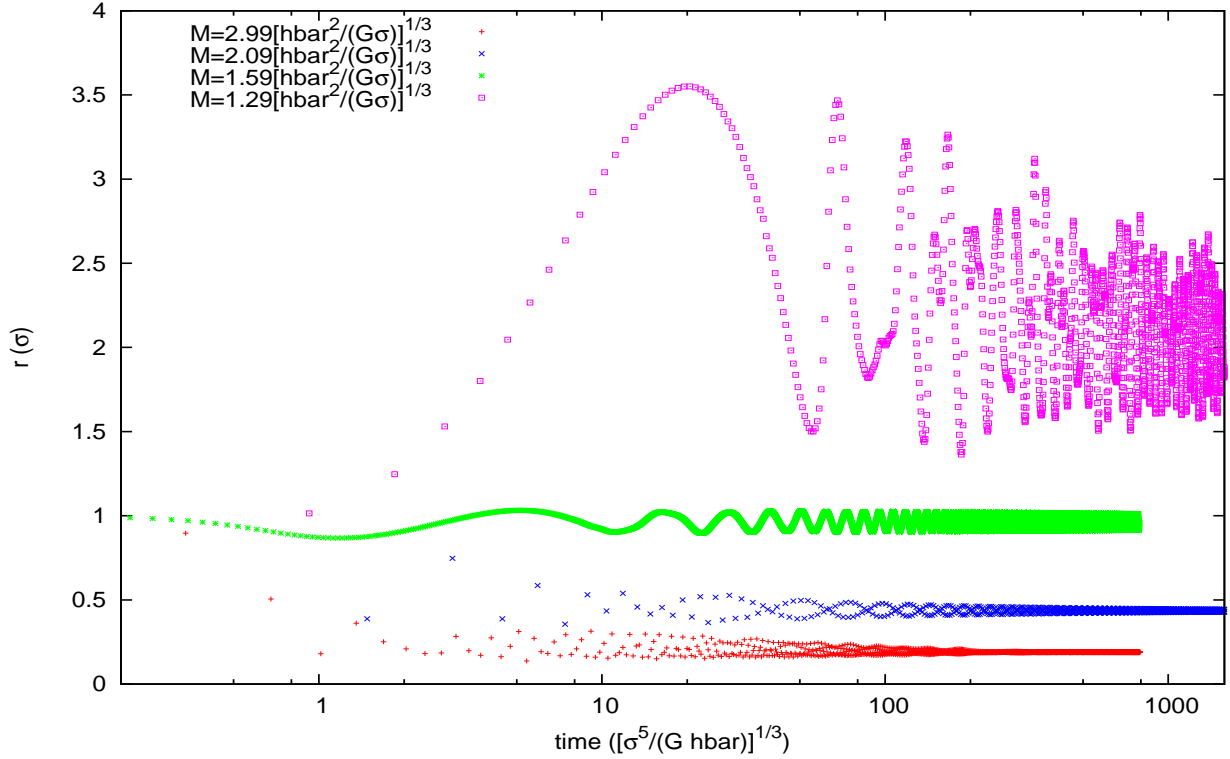}
\caption{Radius of the peak of $r^2|\psi|^2$ versus time for four different
masses.  In each case the position of the peak appears to perform damped 
oscillations about a central value.}
\label{fig:peaktrajectories}
\end{figure}
Fig.~\ref{fig:peaktrajectories} shows several examples of the position of the peak of $r^2|\psi|^2$ oscillating about a central value, with the amplitude of oscillation gradually diminishing over time.
For $1.14(\hbar^2/(G\sigma))^{1/3}<m<1.53(\hbar^2/(G\sigma))^{1/3}$ the wave packet initially expanded, then eventually contracted, before approaching the apparent stationarity.
For $m > 1.53(\hbar^2/(G\sigma))^{1/3}$ most of the wave packet contracted immediately.
Fig.~\ref{fig:r2psi2} shows $r^2|\psi|^2$ late in the evolution for several different masses, after the wave function had become nearly quiescent.
The radius of the peak of $r^2|\psi|^2$ was found to scale roughly with $\hbar^2/(Gm^3)$, as does the width of the MPT groundstate solution found in \cite{Moroz}.
The relationship of these wave packets with the MPT groundstate solution was confirmed by computing $\langle\psi|\hat{H}|\psi\rangle$, which was found to approach $p^3E_0$ (Fig.~\ref{fig:energy}), as in \cite{Harrison}, where $p$ is the fraction of probability that did not disperse away and 
\begin{equation}
E_0=-0.163G^2m^5/\hbar^2
\label{groundstateenergy}
\end{equation}
is the MPT groundstate energy found in \cite{Moroz,Harrison01,Harrison02}.

It is tempting to draw an analogy between the quasistationary final state described above and the groundstate of an 
atom or other conventional potential well.  However, the $p^3$ factor above represents an important distinction.
As first discovered by Harrison et al., a time-dependent Schr\"odinger-Newton solution
can evolve towards what might be described as a fractional groundstate.  
Moroz, Penrose, and Tod originally found that the energy of the lowest energy stationary state, assuming unit normalization, was given by Eq.~(\ref{groundstateenergy}).  However Harrison et al. discovered cases of a wave function that partly dissipates away (as also demonstrated in Fig.~\ref{fig:norm}) while the rest evolves towards $\psi_0'\equiv p^2\psi_0(pr,t)$, where $p$ is the remaining probability and $\psi_0$ represents the MPT groundstate with unit normalization.  (Note that in general if $\psi(r,t)$ is a solution of the Schr\"odinger-Newton equation, then so is $\mu^2\psi(\mu r)$.)  The norm of the wave function $\psi_0'$ can be easily verified to be $p$, and it can similarly be shown $\langle\psi_0'|H|\psi_0'\rangle=p^3E_0$.

\begin{figure} [h]
\includegraphics*[width=25pc,height=40pc,angle=-90]{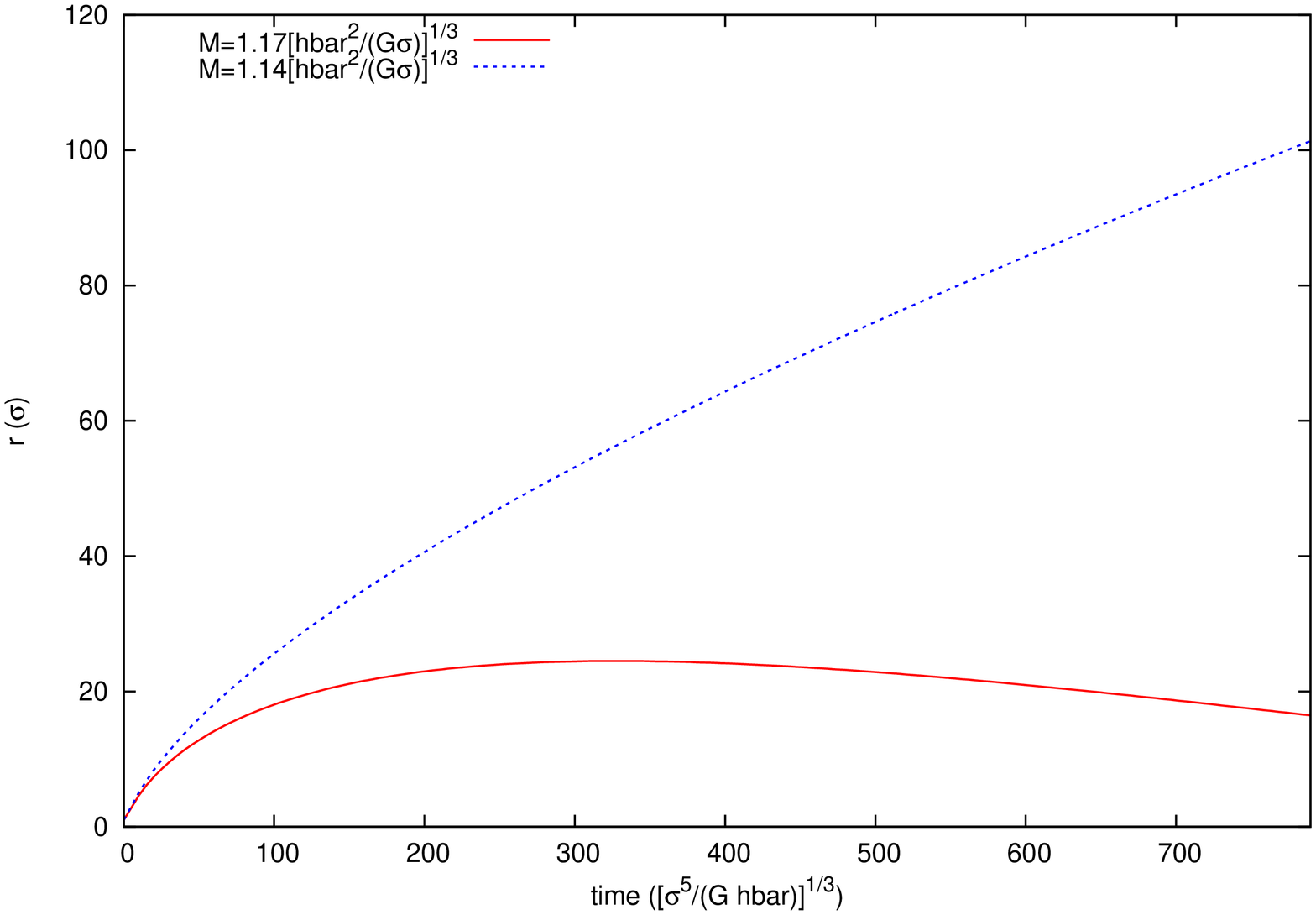}
\caption{The radius of the peak of $r^2|\psi|^2$ vs. t for $m=1.14(\hbar^2/(G\sigma))^{1/3}$ and $m=1.17(\hbar^2/(G\sigma))^{1/3}$.  The latter is bound, 	the former appears to propagate outwards without bound. } \label{fig:escape}
\end{figure}

For $m \leq 1.14(\hbar^2/(G\sigma))^{1/3}$ the entire wave packet appeared to expand indefinitely.  Fig.~\ref{fig:escape} shows the trajectory of the peak of $r^2|\psi|^2$ for $m=1.14(\hbar^2/(G\sigma))^{1/3}$ and
$m=1.17(\hbar^2/(G\sigma))^{1/3}$.  The latter reaches a maximum radius and turns back while the former continues expanding.

\begin{figure} [h]
\includegraphics*[width=25pc,height=40pc,angle=-90]{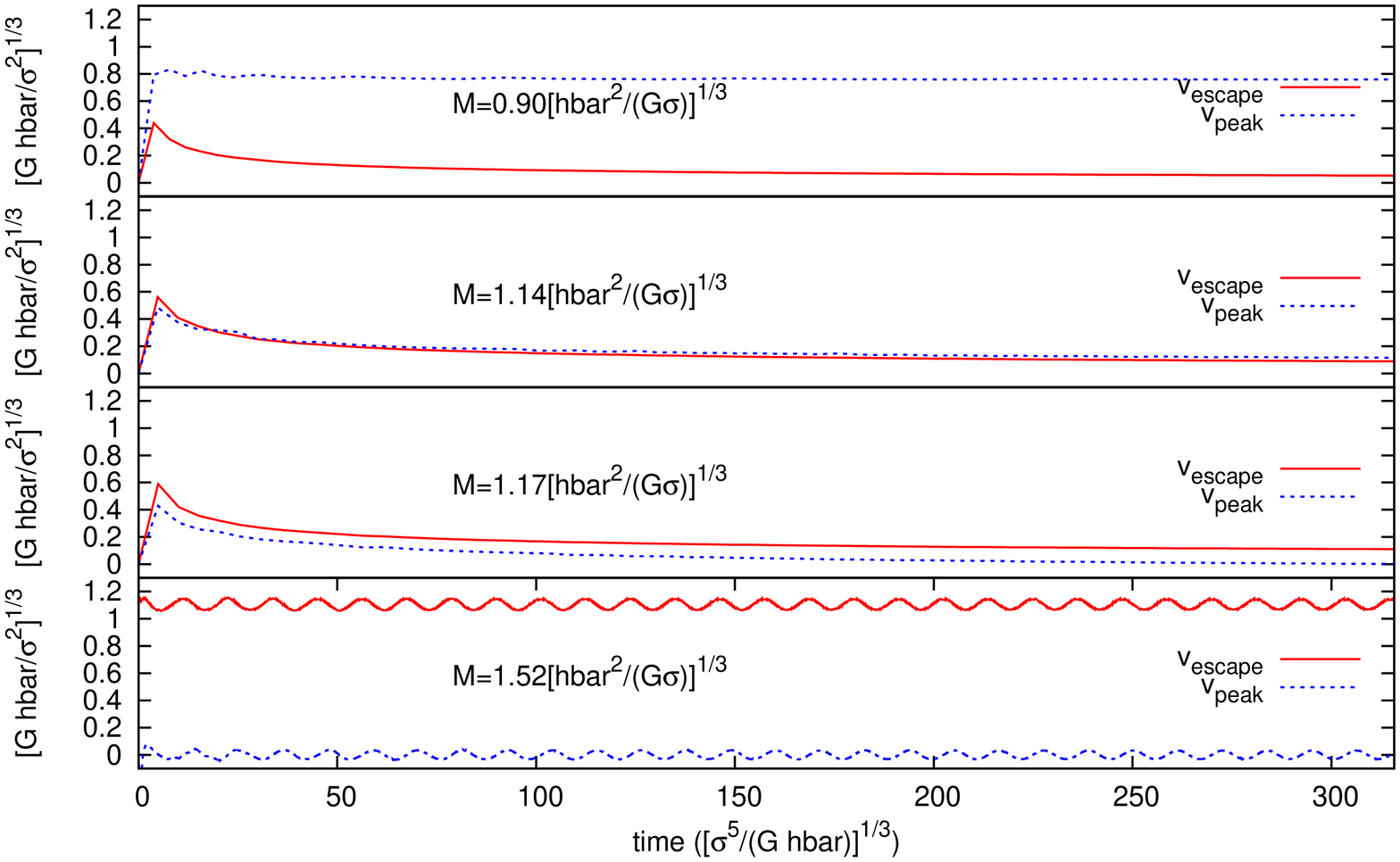}
\caption{The velocity of the peak of $r^2|\psi|^2$ compared with its escape velocity for 	four different masses. }
\label{fig:escapevelocity}
\end{figure}

This threshold is consistent with computation of the Newtonian escape velocity.
Noting from Eq.~(\ref{manyn}) that the effective gravitational mass density is proportional to the probability density, $v_\mathrm{escape}=\sqrt{2Gm_\mathrm{peak}/r_\mathrm{peak}}$, where $m_\mathrm{peak}$ is defined as the effective gravitational mass within $r<r_\mathrm{peak}$ and $r_\mathrm{peak}$ is the radius of the peak.  Fig.~\ref{fig:escapevelocity} compares the velocity of the peak with its escape velocity for four different masses.  The velocity for $m=1.14(\hbar^2/(G\sigma))^{1/3}$ is just above the escape velocity (after some evolution), while the velocity for $m=1.17(\hbar^2/(G\sigma))^{1/3}$ is below the escape velocity, in agreement with Fig.~\ref{fig:escape}.  The velocities for masses below $m=1.14(\hbar^2/(G\sigma))^{1/3}$ were all found to be above the escape velocity and their corresponding trajectories appeared to escape to infinity. The velocities for masses above $m=1.14(\hbar^2/(G\sigma))^{1/3}$ were all found to be below the escape velocity and their corresponding trajectories were clearly bound.

The phenomena described above is also roughly consistent with expectations based on a ``force"-balancing argument of the type given in \cite{Carlipb}.  The initial acceleration due to quantum diffusion for the peak of $r^2|\psi|^2$, where $|\psi|^2$
is Gaussian, and neglecting the gravitational potential, is $\ddot{r}_\mathrm{peak}=\hbar^2/(m^2r_\mathrm{peak}^3)$.  Meanwhile the gravitational acceleration of the peak is $-Gm_\mathrm{peak}/r^2_\mathrm{peak}$.  For a Gaussian wave function, $m_\mathrm{peak}\approx 0.43m$,
so these accelerations are equal and opposite when $m\approx 1.32(\hbar^2/(G\sigma))^{1/3}$.  Of course this argument is only heuristic, but its estimation comes close to the midpoint
in between the ingoing and outgoing solutions.

\section{Discussion}
\label{Discussion}

In summary, for a wave packet of width $\sigma$ and $m>1.14(\hbar^2/(G\sigma))^{1/3}$ Schr\"{o}dinger-Newton solutions evolve towards the MPT groundstate solution.  Since the width of the MPT groundstate solution
scales as $\hbar^2/(Gm^3)$, it can be quite narrow relative to the initial wave packet and the interim evolution resembles the traditional picture of the ``collapse" of the wave function to a Dirac delta function,
albeit through smooth evolution, and with some fraction of the probability dispersing away.
And for $m<1.14(\hbar^2/(G\sigma))^{1/3}$ the wave function seems to spread to infinity, although more slowly than a free particle.

Schr\"{o}dinger-Newton theory makes predictions that deviate significantly from conventional linear quantum mechanics  
when the width of a wave packet is large relative to $\hbar^2/(Gm^3)$.  
Unfortunately, the most massive particle used in a quantum diffraction experiment to date, a ``perfluoroalkylated nanosphere" \cite{Gerlich}, yields $\hbar^2/(Gm^3)\sim 10^{11}$ meters.  
A more reasonable wave packet width of 1 micron, which is close to what can be achieved in a laboratory, would require a mass of
over $10^9$ u to undergo a Schr\"{o}dinger-Newton collapse.
Matter-wave interferometry may not yet be technologically feasible in this
case, but state-of-the-art experiments are expected to reach $10^7$ u for silica spheres laser-cooled in an optical cavity \cite{Romero-Isart}, or even $10^8$ u for gold clusters beamed through diffraction gratings \cite{Nimmrichter}.

\ack
The author thanks S. Carlip for helpful discussion.


\Bibliography{99}
\bibitem{Carlipb} S.\ Carlip, Class.\ Quant.\ Grav.\ 25 (2008) 154010.
\bibitem{Rosenfeld} L.\ Rosenfeld, Nucl.\ Phys.\ 40 (1963) 353.
\bibitem{Moller} C.\ M{\o}ller, in \emph{Les Th{\'e}ories Relativistes
  de la Gravitation}, Colloques Internationaux CNRS 91, edited by
  A.\ Lichnerowicz and M.-A.\ Tonnelat (CNRS, Paris, 1962).
\bibitem{Wald} R.\ Wald, \emph{Quantum field theory in curved spacetime
and black hole thermodynamics}, University of Chicago Press, Chicago, 1994.
\bibitem{Salzman} P.\ J.\ Salzman, ``Investigation of the Time Dependent Schr{\"o}dinger-Newton Equation,'' PhD thesis, University of California at Davis, 2005.
\bibitem{Carlipa} P.\ J.\ Salzman and S.\ Carlip, arXiv:gr-qc/0606120.
\bibitem{Kibble} T.\ W.\ Kibble and S.\ Randjbar-Daemi, J.\ Phys.\ A13
  (1980) 141.
\bibitem{Diosi} L.\ Diosi, Physics Letters 105A (1984) 199.
\bibitem{Adler} S.\ L.\ Adler, J.\ Phys.\ A40 (2007) 755.
\bibitem{Penrose96} R.\ Penrose, General Relativity and Gravitation, 28 (1996) 581.
\bibitem{Penrose98} R.\ Penrose, Phil.\ Trans.\ R.\ Soc.\ 356 (1998) 1927.
\bibitem{Moroz} I.\ M.\ Moroz, R.\ Penrose, and P.\ Tod, Class.\ Quant.\
  Grav.\ 15 (1998) 2733.
\bibitem{Harrison01} R.\ Harrison, ``A numerical study of the Schr\"odinger-Newton equations," PhD thesis, University of Oxford, 2001.
\bibitem{Harrison02} R.\ Harrison, I.\ Moroz, and K.\ Tod, arXiv:math-ph/0208045.
\bibitem{Harrison} R.\ Harrison, I.\ M.\ Moroz, and K.\ P.\ Tod, Nonlinearity
  16 (2003) 101.
\bibitem{Giulini} D.\ Giulini and A.\ Grossardt, arXiv:1105.1921.
\bibitem{Guzman04} F.\ Guzman, L.\ Urena-Lopez, Phys.\ Rev.\ D69 (2004) 124033.
\bibitem{Guzman06} F.\ Guzman, L.\ Urena-Lopez, Astrophys.\ J.\ 645 (2006) 814.
\bibitem{Bernal} A.\ Bernal, F.\ Guzman, Phys.\ Rev.\ D74 (2006) 063504.
\bibitem{Becerril} R. Becerril, F. Guzman, A. Rendon-Romero, and S. Valdez-Alvarado, Revista Mexicana De Fisica E54 (2008) 120. 
\bibitem{Gerlich} S.\ Gerlich, S.\ Eibenberger, M.\ Tomandl, S.\ Nimmrichter, K.\ Hornberger, P.\ Fagan, J.\ T\"{u}xen, M.\ Mayor, and M.\ Arndt, Nature Communications 2 (2011) 186.
\bibitem{Romero-Isart} O.\ Romero-Isart, A.\ Pflancer, F.\ Blaser, R.\ Kaltenbaek, N.\ Kiesel, M.\ Aspelmeyer, J.\ Cirac, Phys.\ Rev.\ Lett.\ 107 (2011) 020405.\bibitem{Nimmrichter} S.\ Nimmrichter, K.\ Hornberger, P.\ Haslinger, M.\ Arndt, Phys. Rev. A83 (2011) 043621.
\endbib
\end{document}